\documentclass[onecolumn,usenatbib,useAMS]{mnras}

\usepackage{amsbsy,amssymb,amsmath,bm,eqnarray}
\usepackage{graphicx,subfigure}
\usepackage{color}
\usepackage{lineno}
\usepackage{changes}

\title[Tidal dynamo in solar-like close binary stars]{Tidal dynamo in solar-like close binary stars}
\author[Xing Wei]{Xing Wei \\ Department of Astronomy, Beijing Normal University, China \\ Email: xingwei@bnu.edu.cn}
\date{}

\begin{document}
\label{firstpage}
\maketitle

\begin{abstract}
Thermal convection is commonly believed to be the energy source of stellar or planetary dynamo. In this short paper we provide another possibility, namely large-scale tidal flow. In close binary stars, say, solar-like stars with orbital period at 2 or 3 days, large-scale tidal flow is comparable to or even stronger than convective flow, and it can induce magnetic dynamo action. Based on dynamo equation and tidal theory we estimate the magnetic energy induced by large-scale tidal flow, which is proportional to the cube of orbital frequency. Our estimation can be tested by the future spectropolarimetric observations and numerical simulations for close binary stars.
\end{abstract}

\section{Motivation}
Magnetic dynamo in star or planet can be induced by various mechanisms, e.g. thermal convection \citep{christensen2010, jones2010, jones2011}, precession \citep{tilgner2005, wei2016a}, collision \citep{wei2014}, and mechanical stirring \citep{stevenson2011}. Tide is believed to be much weaker than convection so that the tidally induced dynamo is not widely studied. Some researchers tried this mechanism for Mars \citep{arkani2009} and exoplanets \citep{knapp2017}. \citet{cebron2014} performed numerical calculations for this mechanism to find that the elliptical instability in tidal flow can indeed drive magnetic dynamo. \citet{vidal2018} numerically studied the tidal dynamo in a radiative star where the elliptical instability plays a key role. The effect of elliptical instability on magnetic field was also studied in \citet{cebron2012, barker2014, reddy2018, vidal2019}. Tidal flow can be decomposed into two parts, namely the equilibrium tide in a quasi-hydrostatic balance and the dynamical tide of fluid waves, e.g. internal gravity waves or inertial waves \citep{ogilvie2014}. The elliptical instability is essentially induced by small-scale inertial waves, namely a type of dynamical tide, and this instability obtains energy from the large-scale equilibrium tide in the absence of convection. In this paper, we will estimate the magnetic energy induced by tidal flow alone (i.e. without convection), and more specifically by the equilibrium tide, and apply it to close binary solar-like stars with orbital period at 2 or 3 days \citep{mathieu2005}. It should be noted that our estimation is for the overall energy budget but not the specific flow pattern. We will estimate how large the fraction of tidal energy is converted to magnetic energy.

\section{Estimation}
In the first place we estimate the tidal flow in close binary stars, which refers to the equilibrium tide only in the following, and compare it to convective flow. We denote primary (on which tide is raised) by subscript 1 and secondary (by which tide is raised) by subscript 2, gravitational constant by $G$, mass by $M$, radius by $R$, and surface gravity by $g$. We then give the estimations of tidal potential per unit mass $\Psi\approx GM_2R_1^2/a^3$ where $a$ is orbital separation, tidal force per unit mass $f\approx\Psi/R_1\approx GM_2R_1/a^3$, tidal deformation $\xi\approx\Psi/g_1\approx(M_2/M_1)(R_1^4/a^3)$, and tidal flow $u\approx\omega\xi\approx\omega(M_2/M_1)(R_1^4/a^3)$ where $\omega$ is orbital frequency (strictly speaking it is tidal frequency but approximately orbital frequency when stars rotate slowly not at their young age). Suppose that the binaries are solar-like stars with solar mass and radius. With orbital period at 3 days tidal flow $u\approx 30$ m/s, and with orbital period at 2 days tidal flow $u\approx 90$ m/s. On the other hand, we use the mixing length theory to estimate the convective velocity $v$ \citep[e.g.][]{maeder2009}. This theory states that the work done by buoyancy force is converted to kinetic energy, $\delta\rho g l\approx\rho v^2$ where $\rho$ is density, $\delta\rho$ is density deviation from surroundings and $l$ is the mixing length on which the turbulent momentum transport completes. The mixing length is proportional to the pressure scale height, $l\approx\alpha p/(dp/dr)\approx\alpha p/(\rho g)\approx\alpha(\mathcal R/\mu)(T/g)\approx c_pT/g$ where $\alpha$ is a model coefficient between 1 and 2, and the hydrostatic balance $dp/dr\approx\rho g$, the equation of state for ideal gas $p=\rho\mathcal{R}T/\mu$ ($\mathcal{R}$ being gas constant, $T$ temperature, and $\mu$ molecular weight) and specific heat capacity at constant pressure $c_p=2.5\mathcal R/\mu$ in convection zone are employed. Inserting $l\approx c_pT/g$ into $\delta\rho g l\approx\rho v^2$ we are led to $v^2\approx(\delta\rho/\rho)c_pT$. With $\delta\rho/\rho\approx\delta T/T$ where $\delta T$ is temperature deviation from surroundings we obtain $v^2\approx c_p\delta T$. Introducing heat flux $q=\rho vc_p\delta T\approx\rho v^3$ \citep{spiegel1963}, we obtain $v\approx(q/\rho)^{1/3}$, which has been already validated by numerical simulations \citep{chan1996, cai2014}. Putting the solar values, we can estimate the volume-averaged convective velocity $v\approx 30$ m/s. The asteroseismology shows that the solar convective velocity $v$ is about 10 m/s or lower \citep{hanasoge2012}. Therefore, with orbital period at 3 days tidal flow is comparable to convective flow, and with orbital period at 2 days tidal flow is even stronger than convective flow. Since convective flow can induce magnetic dynamo action, such a strong tidal flow should also have this capability. Next we estimate the tidal dynamo.

We write down the total energy equation of tidal dynamo in primary,
\begin{equation}\label{energy}
\frac{\partial}{\partial t}\left(\frac{\rho u^2}{2}+\frac{B^2}{2\mu}\right)=-\bm\nabla\cdot\bm A+\rho\bm f\cdot\bm u-D_\nu-\frac{J^2}{\sigma},
\end{equation}
where the left-hand-side is the rate of the total energy consisting of kinetic energy and magnetic energy, $\bm u$ is tidal flow, $\bm B$ is magnetic field ($\mu$ is magnetic permeability and different from molecular weight as used in the last paragraph), $\bm A$ is the total flux consisting of kinetic energy flux $(\rho u^2/2)\bm u$, pressure energy flux $p\bm u$ and Poynting flux $(\bm E\times\bm B)/\mu$ ($\bm E$ being electric field), $\bm f$ is tidal force per unit mass, $D_\nu$ is viscous dissipation, and the last term is Ohmic dissipation ($J$ being electric current and $\sigma$ electrical conductivity). We calculate the volume integral over the convection zone for dynamo action. The total energy is statistically steady so that the left-hand-side vanishes. The net flux almost vanishes, because the energy carried by stellar wind away from stellar surface is tiny compared to internal energy. Viscous dissipation is much smaller than Ohmic dissipation \citep{wei2016b, wei2018, ogilvie2018, astoul2019}. Thus, the two terms are left to balance each other
\begin{equation}\label{balance1}
\langle J^2/\sigma\rangle\approx\epsilon\langle\rho fu\rangle
\end{equation}
where brackets denote the volume average in primary's convection zone and the parameter $\epsilon$ arises from the orientation between tidal force $\bm f$ and tidal flow $\bm u$, namely the dot product $\bm f\cdot\bm u$ in \eqref{energy}. Although we cannot tell exactly the orientation of the two vectors since the tidal flow pattern is complex in the convection zone, we can measure this effect by the parameter $\epsilon$ which can be fortunately estimated. Equation \eqref{balance1} states that a fraction $\epsilon$ of the power of tidal flow eventually goes to Ohmic dissipation. Integrating \eqref{balance1} for one orbital cycle, we will find that the left-hand-side is the tidal dissipation during one orbit and the integral of $\langle\rho fu\rangle$ on the right-hand-side is the energy stored in tide. Thus, according the definition of tidal $Q$ number \citep{goldreich1966}, the coefficient $\epsilon$ is at the order of the inverse of tidal $Q$ number, i.e. $\epsilon\approx Q^{-1}$. In the standard tidal theory (ch8 in \citet{tides-book}), $Q^{-1}$ is the tidal angle and it can be modelled as $\omega/(\tau_dGM_1/R_1^3)$ where the dissipation timescale $\tau_d$ in our situation is no longer viscous dissipation timescale $R_1^2/\nu_t$ ($\nu_t$ being turbulent viscosity) but Ohmic dissipation timescale $R_1^2/\eta_t$ ($\eta_t$ being turbulent magnetic diffusivity). Therefore, we find the estimation for the parameter $\epsilon$
\begin{equation}\label{epsilon}
\epsilon\approx Q^{-1}\approx \frac{\omega\eta_t}{GM_1/R_1}.
\end{equation}
The observational constraints \citep{choudhuri2007} and numerical simulations \citep{brandenburg2003, brandenburg2020} have already shown that in a solar-like star turbulent magnetic diffusivity $\eta_t$ and turbulent viscosity $\nu_t\approx vl$ are at the same order of magnitude, namely $10^{12}~{\rm cm^2/s}$.

Here we need to clarify the effect of rotation on magnetic dynamo. \citet{christensen2006} and \citet{christensen2009} derived the scaling law for the magnetic energy in the convection-driven dynamo which is independent of rotation. \citet{davidson2013} derived the scaling law for the planetary dynamo which depends on rotation. The observations show that stellar magnetic field indeed depends on rotation \citep{wright2011, lehtinen2020}. Recently, \citet{wei2022} studied the convection-driven dynamo and pointed out that magnetic energy is independent of rotation at Rossby number $Ro>1$ (the ratio of inertial force to Coriolis force) but scales as $Ro^{-2}$ at $Ro<1$ and saturates at very small $Ro$. When we consider the magnetic energy equation, the Coriolis force does not enter the energy equation since it is perpendicular to fluid velocity. That is the reason why the scaling law in \citet{christensen2006} and \citet{christensen2009} is independent of rotation. However, the rapidly rotating turbulence is anisotropic and \citet{wei2022} found that the mixing length theory used for the estimations of magnetic energy breaks at fast rotation, because the force balance is no longer between buoyancy force and inertial force but between buoyancy force and Coriolis force, and consequently rotation enters the scaling law for magnetic energy as $\sim Ro^{-2}$ (the saturation of magnetic energy is caused by the growth limit of columnar turbulent eddies, i.e. the depth of convection zone). Back to the tidal dynamo studied in this paper, the tidal flow $u\approx\omega(M_2/M_1)(R_1^4/a^3)$ is the equilibrium tide on the large scale, namely primary's radius $R_1$ which is irrelevant to the convection mixing length. Therefore, for the dynamo driven by the large-scale equilibrium tide, rotation is unimportant. Rotation influences the small-scale dynamical tide of inertial waves, i.e. the length scale of inertial waves scales as Ekman number$^{0.1\sim 1}$ \citep{kerswell1995, rieutord2001, rieutord2018, favier2014}, and the dynamo driven by small-scale inertial waves is plausibly related to rotation, but this type of dynamo will not be studied in this paper.

Next, by introducing the magnetic length scale $l_B$, we use Ampere's law to estimate the Ohmic dissipation $J^2/\sigma\approx(B/\mu l_B)^2/\sigma$. The length scale $l_B$ can be further estimated by the magnetic induction equation. We assume that magnetic induction takes effect on the length scale of tidal flow $\bm u$, namely primary's radius $R_1$ (recall that the stretching term for field amplification $\bm B\cdot\bm\nabla\bm u$ and the tidal flow $u\approx\omega(M_2/M_1)(R_1^4/a^3)$), and magnetic diffusion takes effect on small scale $l_B$, such that we find $l_B\approx(\eta_t R_1/u)^{1/2}$. It should be noted that in this analysis we do not assume that dynamo is on large scale (indeed there exist many small-scale dynamos) but we address that the magnetic induction works on large scale and magnetic diffusion works on small scale. It should be also noted that this magnetic diffusion length scale $l_B$ is different from the large length scale $R_1$ used for Ohmic dissipation timescale $\tau_d$ to model tidal $Q$ number. Equilibrium tide is on the large length scale $R_1$ and its dissipation takes effect on this large length scale. Consequently, the tidal $Q$ number which measures the tidal dissipation efficiency is reasonably modelled with $R_1$. Dynamo has a different physical mechanism that is the competition between magnetic induction on the large length scale $R_1$ and magnetic diffusion on the small length scale $l_B$. This two-scale analysis is widely used in diffusive systems, e.g. boundary layer where viscosity determines the thickness \citep{landau1987}. Otherwise, if we admitted that the magnetic induction and the magnetic diffusion work on the same length scale then we would find that the magnetic Reynolds number is of order of unity at which dynamo cannot be driven. Inserting the estimation of $l_B$ into the expression of Ohmic dissipation, we readily obtain $J^2/\sigma\approx(B^2/\mu)(u/R_1)$. Inserting this estimation into \eqref{balance1}, we obtain the estimation of magnetic energy
\begin{equation}\label{balance2}
\langle B^2/\mu\rangle\approx\epsilon\langle\rho fR_1\rangle
\end{equation}
which states that a fraction $\epsilon$ of the work done by the large-scale tidal flow generates magnetic energy, i.e. the essence of tidal dynamo. At the last step, we insert the estimations of mean density $\bar\rho\approx M_1/R_1^3$ and tidal force per unit mass $f\approx GM_2R_1/a^3$ into \eqref{balance2} to arrive at
\begin{equation}\label{estimation1}
\langle B^2/\mu\rangle\approx\epsilon (GM_1M_2)/(R_1a^3)\approx\epsilon M\omega^2/R_1
\end{equation}
where Kepler's third law $\omega^2=G(M_1+M_2)/a^3$ is used and $M$ is the reduced mass $M=(M_1M_2)/(M_1+M_2)$.

In the above crude estimation \eqref{estimation1} we did not consider the density profile on the right-hand-side of \eqref{balance2}, which will bring a structure factor $\beta$ in front of the right-hand-side of \eqref{estimation1}, 
\begin{equation}\label{beta}
\beta=\frac{3}{1-\alpha^3}\int_\alpha^1(\rho/\bar\rho)(r/R_1)^3d(r/R_1)
\end{equation}
where the base of convection zone in solar-like stars is located at $r/R_1=\alpha\approx 0.7$. Note that in the derivation of $\beta$ \eqref{beta} we use the more exact expression of tidal force $f\approx GM_2r/a^3$ instead of its crude estimation $f\approx GM_2R_1/a^3$. Therefore, we eventually obtain the estimation of magnetic energy
\begin{equation}\label{estimation2}
\langle B^2/\mu\rangle\approx\beta\epsilon M\omega^2/R_1.
\end{equation} 
With the aid of the MESA code \citep{Paxton2011} to calculate the solar structure, we integrate \eqref{beta} to obtain $\beta\approx 0.03$. 

With the solar values and the orbital period at 2 days, the estimation \eqref{estimation2} together with \eqref{epsilon} gives the volume-averaged magnetic field $\langle B\rangle\approx 500$ Gauss. In the stellar interior, the azimuthal field can be locally very strong near tachocline due to the strong shear, for example, it can reach a few ten Tesla \citep{charbonneau2013}. But the surface field cannot be so strong and we can make a simple estimation. Suppose magnetic field is dipolar and scales as $B_s(r/R)^{-3}$ where $B_s$ is surface field. We take the volume integral to calculate $\langle B\rangle=(1/V)\int B_s(r/R)^{-3}dV=-3\ln\alpha/(1-\alpha^3)B_s\approx 2B_s$. In \citet{christensen2009}, this ratio $\langle B\rangle/B_s$ is taken to be 3.5. Therefore, the surface field is roughly 200 Gauss.

\section{Summary and discussions}
In this paper we derived the estimation of magnetic energy \eqref{estimation2} in dynamo induced by tidal flow. Combining \eqref{epsilon} and \eqref{estimation2} we are led to
\begin{equation}
\langle B^2/\mu\rangle\approx\beta(M/M_1)\eta_t\omega^3/G.
\end{equation} 
This formula shows that magnetic energy scales as $\omega^3$. On the other hand, the theoretical studies show that turbulent viscosity $\nu_t$ can be suppressed by fast tide to follow the scaling law $\omega^{-1}$ \citep{zahn1977} or $\omega^{-2}$ \citep{goldreich1977} or $\omega^{-5/3}$ \citep{goodman1997}. The recent studies support the $\omega^{-2}$ suppression law \citep{ogilvie2012, duguid2020, vidal2020}. Since turbulent viscosity $\nu_t$ and turbulent magnetic diffusivity $\eta_t$ are caused by the same process, namely the turbulent transport, they possibly follow the similar suppression law. If we admit this point then the different suppression laws for turbulent magnetic diffusivity $\eta_t$ versus tidal frequency $\omega$ will yield the different scaling laws for magnetic energy $\langle B^2/\mu\rangle$ versus tidal frequency $\omega$. The spectropolarimetric observations and the numerical simulations in the future can justify or deny our estimation for magnetic energy, and moreover, justify the suppression law for turbulent magnetic diffusivity.

\section*{Acknowledgments}
This work is supported by National Natural Science Foundation of China (11872246, 12041301) and Beijing Natural Science Foundation (1202015).

\section*{Data availability}
The data underlying this article are available in the article.

\bibliographystyle{mnras}
\bibliography{paper}

\label{lastpage}
\end{document}